\documentclass{emulateapj}
\usepackage{apjfonts}

\newcommand{\timav}{\langle \dot M \rangle}

\shorttitle{Accreting White Dwarf Surface Temperatures} 

\begin{document}

\submitted{Accepted by ApJL}
\title{Measuring White Dwarf Accretion Rates via
their Effective Temperatures} 

\author{Dean M. Townsley}
\affil{Department of Physics\\
Broida Hall, University of California, Santa Barbara, CA 93106;
townsley@physics.ucsb.edu}
\author{Lars Bildsten}
\affil{Kavli Institute for Theoretical Physics and Department of Physics\\
Kohn Hall, University of California, Santa Barbara, CA 93106;
bildsten@kitp.ucsb.edu}

\begin{abstract}

Our previous theoretical study of the impact of an accreting envelope
on the thermal state of an underlying white dwarf (WD) has yielded
equilibrium core temperatures, classical nova ignition masses and
thermal luminosities for WDs accreting at time averaged rates of
$\timav=10^{-11}-10^{-8} M_\odot \ {\rm yr^{-1}}$. These $\timav$'s
are appropriate to WDs in cataclysmic variables (CVs) of $P_{\rm
orb}\lesssim 7$ hr, many of which accrete sporadically as Dwarf
Novae. Approximately thirty nonmagnetic
Dwarf Novae have been observed in quiescence, when the accretion rate
is low enough for spectral detection of the WD photosphere, and a
measurement of $T_{\rm eff}$. We use our theoretical work to translate
the measured $T_{\rm eff}$'s into local time-averaged accretion
rates, confirming the factor of ten drop in $\timav$ predicted for 
 CV's as they transit the period gap. For DN below the period gap, we
show that if $\timav$ is
that given by gravitational radiation losses alone, then the WD masses
are $>0.8M_\odot$. An alternative conclusion is that the masses are
closer to $0.6M_\odot$ and $\timav$ is 3--4 times larger than that
expected from gravitational radiation losses. In either case, it is
very plausible that a subset of CVs with $P_{\rm orb}<2$ hours will
have $T_{\rm eff}$'s low enough for them to become non-radial
pulsators, as discovered by van Zyl and collaborators in GW Lib.

\end{abstract}


\defcitealias{TownBild03}{TB}

\section{Introduction}

Cataclysmic variables (CVs; \citealt{Warn95}) are formed when the
low-mass stellar companion of a WD, exposed during a common envelope
event, finally (on timescales of $0.1-10$ Gyr) fills the Roche lobe
as a result of long term angular momentum losses.  The WD will
cool during this time; a $0.20 M_\odot$ He WD would have
$T_c=3.3\times 10^6 \ {\rm K} $ at 4 Gyr \citep{AlthBenv97}, whereas a
$0.6M_\odot$ C/O WD would have $T_c=2.5\times 10^6 {\rm K}$ in 4 Gyr
\citep{Salaetal00}. These WDs have effective temperatures $T_{\rm
eff}\approx 4500-5000$ K just before mass transfer starts.
However, once the Roche lobe is filled, the WD accretes material at
$\timav\approx 10^{-11}-10^{-8} M_\odot \ {\rm yr^{-1}}$ (e.g.\
Howell, Nelson, \& Rappaport 2001) and can be reheated \citep{Sion91}
to higher $T_{\rm eff}$'s.

 Dwarf Novae (DN) are the subset of CVs with low time-averaged rates
$\timav<10^{-9}M_\odot \ {\rm yr}^{-1}$ and thermally unstable
accretion disks. The transfer of matter onto the WD occurs in
outbursts that last a few days to a week once every month to year (or
even longer in some systems). Most DN have orbital periods $P_{\rm
orb} < 2$ hours, below the ``period gap'', with fewer above the gap
\citep[see][]{Shaf92}. During accretion disk quiescence, the $\dot M$
onto the WD is often low enough that the system's UV (and sometimes
optical) emission is dominated by light from the WD surface, allowing
for a measurement of the WD $T_{\rm eff}$, nearly all of which exceed
$10,000$ K \citep{Sion99}. Thus, the WD is hotter than expected for its
age, providing evidence of the thermal impact of prolonged accretion
on the WD \citep{Sion95}.
\citet{TownBild03} (hereafter \citetalias{TownBild03})
have calculated $T_{\rm eff}$ and its dependence on $\timav$, the WD
mass, $M$, and core temperature, $T_c$. In this paper we now use that
work to constrain these parameters from the measured $T_{\rm eff}$'s.

  The gravitational energy released when a particle falls from a large
distance to the stellar surface ($GM/R$) is deposited near the
photosphere and is rapidly radiated away. This energy does not
penetrate inwards with the inflowing material, as the time it takes
the fluid to move inward is much longer than the time it takes for
heat to escape. This eliminates the outer boundary condition and
instead points to the importance of energy release deep in the
accreting H/He envelope due to both gravitational energy release and a
low level of nuclear ``simmering'' \citepalias{TownBild03}.
We begin in \S
\ref{sec:review} by reviewing our work and showing that the best
constrained quantity from a measured $T_{\rm eff}$ is $\langle \dot m
\rangle \equiv \timav/4\pi R^2$. At very low $\timav$'s,  $T_{\rm eff}$
also depends on $T_c$. However, in this regime, we can calculate $T_c$
self consistently \citepalias{TownBild03},
removing it from consideration.

In \S \ref{sec:measurements} the $\langle \dot m\rangle$'s implied
by the available measurements are presented and compared to CV
evolutionary scenarios. A detailed comparison is presented for DN with
$P_{\rm orb}\lesssim 2$ hours, showing that the observed $T_{\rm
eff}$'s imply that either $M>0.6M_\odot$ or $\timav$ is larger 
than implied by gravitational radiation losses alone.
We close in \S \ref{sec:conclusion} 
with a discussion of future work, especially the seismology
of accreting WDs.

\section{The Relation of $T_{\rm eff}$ to $\timav$}
\label{sec:review}

\begin{figure*}
\plotone{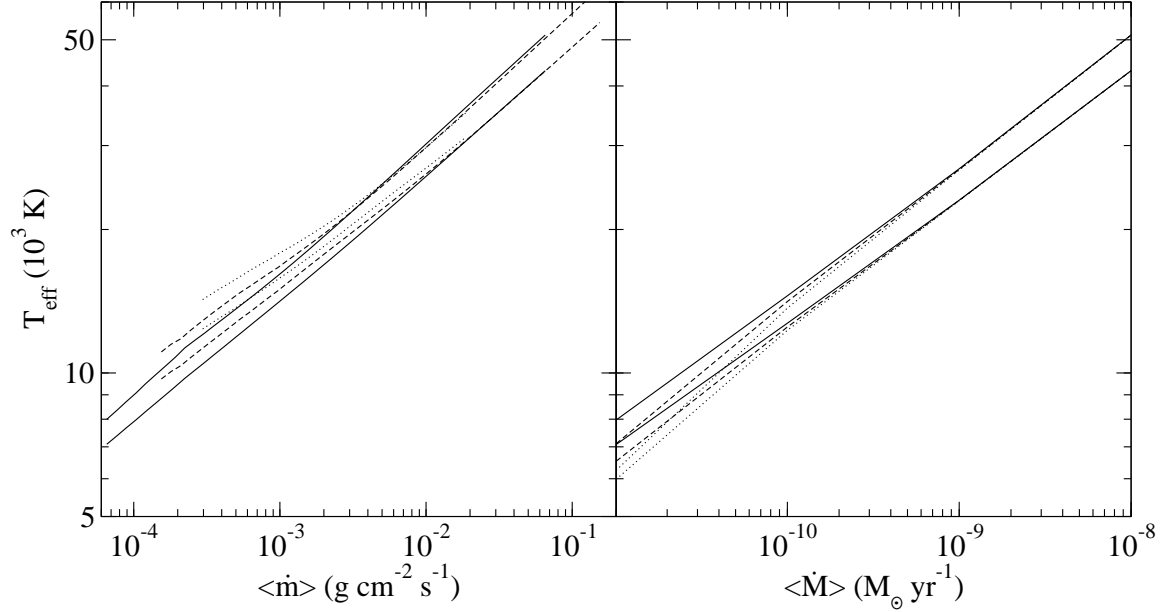}
\caption{
\label{fig:Teff-mdot}
Left Panel:
Predicted range of $T_{\rm eff}$ (between lines) for $0.05M_{\rm
ign} < M_{\rm acc} < 0.95M_{\rm ign}$ at $M=0.6M_\odot$ (solid lines),
$1.0M_\odot$ (dashed lines), and $1.2M_\odot$ (dotted lines),
Without knowledge of $M$, $\langle \dot m\rangle$ is still fairly well
constrained from $T_{\rm eff}$.
\label{fig:Tcdemo}
Right Panel:
Dependence of $T_{\rm eff}$ range (as in left panel)
on $\timav$ for $T_c = T_{c,\rm eq}$
(solid lines), $0.75T_{c,\rm eq}$ (dashed lines), and $0.5T_{c,\rm
eq}$ (dotted lines), at $M=0.6M_\odot$.  This observable does not
depend strongly on $T_c$ except at the lowest $\timav$'s, and
therefore is relatively insensitive to the CVs  evolution.}
\vspace{-0.3in}
\end{figure*}

   The high quality UV spectra of quiescent DN from the STIS 
instrument on \emph{Hubble
Space Telescope}, e.g.\ \citet{HoweGansetal02},
\citet{SzkoGansetal02} and many of the references in the table
appearing in \citet{WintSion03}, yield accurate measurements of
$T_{\rm eff}$. However, the lack of accurate distance information
prohibits the measurement of the WD radius, and hence mass,
motivating the identification of a physical parameter which is best
constrained by $T_{\rm eff}$ alone.
The intent of this section is to
demonstrate that, without knowledge of $R$ or $M$, the best
constrained parameter is the accretion rate per unit area,
$\langle \dot m\rangle$.

Our previous work \citepalias{TownBild03} presented a detailed discussion
of the impact of accretion on the thermal structure of a WD.  For this
paper we are primarily interested in the predictions for the surface
luminosity, $L(M,\timav,T_c)$, for which we showed that the strongest
dependence is $L\simeq \timav T_c/\mu m_p= 4\pi R^2 \sigma_{SB}T_{\rm
eff}^4$ and thus $T_{\rm
eff}^4\propto \timav/4\pi R^2=\langle\dot m\rangle$, even for
different masses.
The prime
variance in this simple mapping is from the change in $T_{\rm eff}$ as
the mass of the accumulated layer, $M_{\rm acc}$, increases between
classical novae (CN). In discussing our theoretical results we use the
$T_{\rm eff}$ range for $0.05M_{\rm ign}\le M_{\rm acc} \le 0.95
M_{\rm ign}$, which represents where an observed WD is most likely to
be found. We set $X_{^3\rm He}=0.001$ throughout as the difference
between this and similar predictions for $X_{^3\rm He}=0.005$ is less
than the uncertainty due to the unknown $M_{\rm acc}$.  Figure
\ref{fig:Teff-mdot} shows the range of $T_{\rm eff}$ traversed as a
function of $\langle \dot m\rangle$ for $M=0.6$-$1.2M_\odot$,
showing that $T_{\rm eff}$ provides a reasonable
constraint on $\langle \dot m\rangle$ even when $M$ is not known.
A $T_{\rm eff} = 15,000$ K implies that $0.4\times 10^{-3} \le
\langle\dot m\rangle \le 1.3\times10^{-3}$ g cm$^{-2}$ s$^{-1}$,
whereas
a $T_{\rm eff} = 30,000$ K implies that $1.0\times 10^{-2} \le
\langle\dot m\rangle \le 1.7\times10^{-2}$ g cm$^{-2}$ s$^{-1}$.

 The insensitivity of $L$ to $M$ is the strongest qualitative
difference between our calculation and Sion's (1995) estimate of
$L\simeq0.2\timav GM/R$, made from the steady state models of
\citet{Iben82}. While Iben's (1982) models are steady state in the
sense of $L_{\rm core}=0$, they were for high accretion rates, $\timav
=1.5\times 10^{-8}M_\odot$ yr$^{-1}$ for the best discussed models,
with one at $\timav = 1.5\times 10^{-9}M_\odot$ yr$^{-1}$, and use
steadily burning shells. Such a steady state is inappropriate at the
low $\timav$'s discussed here, where the burning is always unstable.
Iben's (1982) models are also qualitatively very different from ours:
in our models compressional energy released in the accreted H/He shell
provides the outgoing luminosity and helps to establish the
equilibrium configuration of the core, whereas in Iben's (1982)
models, the compressional heating term is entirely in the core, a
contribution which is likely small (see
Appendix A of \citetalias{TownBild03}). 
The results of TB are complementary to \citet{GodoSion02},
which focuses on the
response of the WD to the short-timescale
$\dot M$ variations during DN outbursts.

 The last parameter dependence to explore is $T_c$, the WD core
temperature. As discussed in \citetalias{TownBild03}, DN below the period
gap have adequate time to reach an equilibrium core temperature,
$T_{c, \rm eq}$, that depends on $\timav$ and $M$.  However, DN above
the period gap have not been accreting long enough for $T_c$ to reach
$T_{c, \rm eq}$. Figure \ref{fig:Tcdemo} shows how the traversed
$T_{\rm eff}$ range depends on $T_c$ for a 
$M=0.6M_\odot$ WD. The curves show $T_{\rm eff}$
for $T_c=T_{c,\rm eq}$, $0.75T_{c,\rm eq}$ and $0.5 T_{c,\rm eq}$. Due
to a strong core/envelope decoupling for $\timav >10^{-10}M_\odot$
yr$^{-1}$ \citepalias{TownBild03}, the $T_{\rm eff}$ range is
nearly independent of $T_c$.  For lower $\timav$'s, the WD 
core temperature should be close to 
the equilibrium value, allowing us 
to use $T_{c,\rm eq}$ as representative when finding 
$\langle \dot m\rangle$.

\begin{figure}
\plotone{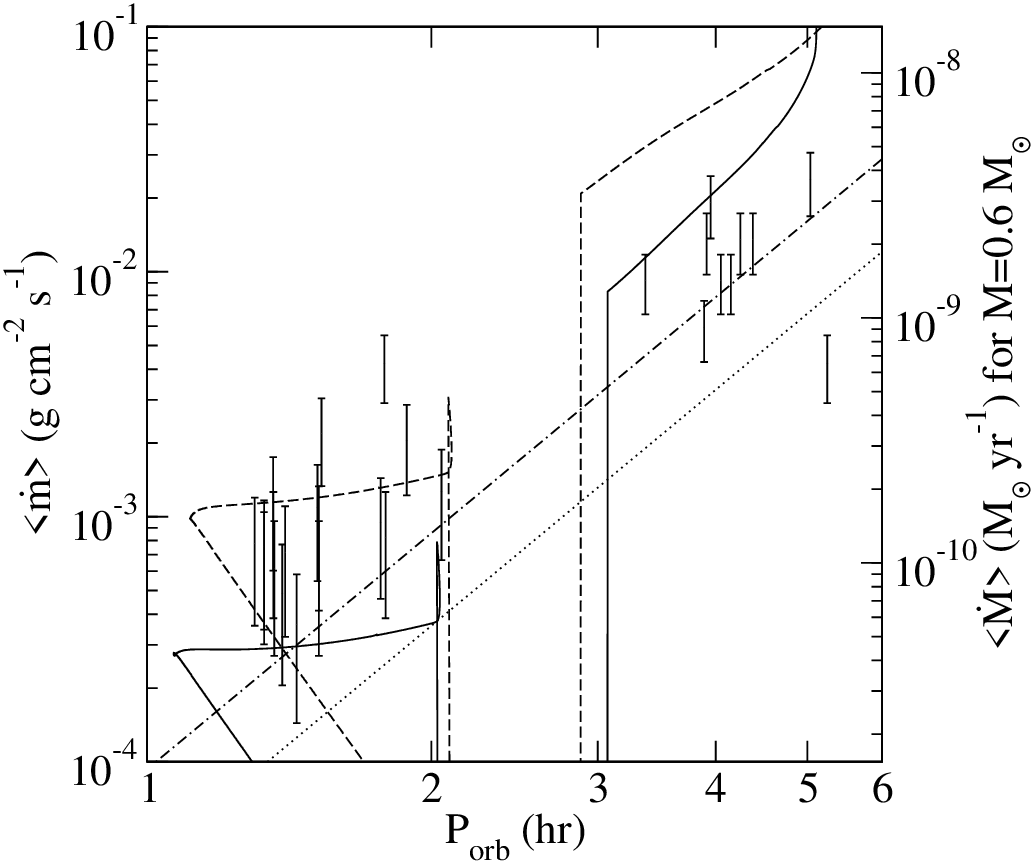}
\caption{
\label{fig:mdot-Porb_wide}
Values for the time averaged accretion rate per WD surface area, $\langle
\dot m\rangle\equiv \timav/4\pi R^2$, derived from the $T_{\rm eff}$
measurements in the table appearing in \cite{WintSion03}.
The ranges indicated for each measurement are those allowed for $0.05 M_{\rm
ign} < M_{\rm acc} < 0.95M_{\rm ign}$ and $0.6M_\odot <M< 1.2M_\odot$.
The curves show the $\langle \dot m \rangle$ predicted by
\citet*{Howeetal01} for $M=0.6M_\odot$ (solid line) and $M=1.1M_\odot$
(dashed line). Patterson's (1984) deduced relation from
CV observations is shown by the dotted line for 
$M=0.6M_\odot$ the dot-dashed line for $1.0M_\odot$. 
The right hand scale gives $\timav_{0.6}$, the corresponding accretion
rate if $R$ is that for $M=0.6M_\odot$.
At the same $\langle \dot m\rangle$, $\timav_{1.0} =
0.4\timav_{0.6}$.
}
\vspace{-0.3in}
\end{figure}

\section{Inferring Accretion Rates from $T_{\rm eff}$ Measurements}

\label{sec:measurements}

Figure \ref{fig:mdot-Porb_wide} shows the $\langle\dot m\rangle$'s
inferred from the measured $T_{\rm eff}$'s tabulated in
\cite{WintSion03}, with the following modifications: for WX Cet and VY
Aqr we use the single-temperature fitted values, for CU Vel we correct
the misquoted value, for AL Com we use a measurement longer
after superoutburst \citep{Szkoconf02}, and we add GW Lib
\citep{SzkoGWLib02} and DW UMa \citep{Szkoconf02}. This observational
$\langle\dot m\rangle$-$P_{\rm orb}$ relation shows clear evidence for
a drop in $\langle\dot m\rangle$ below the period gap.

The relationship between $P_{\rm orb}$ and $\timav$ is still a very
active area of theoretical inquiry and one that we hope our work can
illuminate. The expectations from ``standard'' CV evolution
\citep{Howeetal01} for $M=0.6M_\odot$ (solid line) and $1.1M_\odot$
(dashed line) are shown in Figure 2. In this disrupted magnetic
braking scenario, $\timav$ is set by magnetic braking above the period
gap and by gravitational radiation below the period gap. Our deduced
$\langle\dot m\rangle$'s are lower than the expected values above the
period gap. It is important that we are inferring the long-term
$\timav$, averaged over the thermal time of the radiative
(nondegenerate) layer $\sim c_P T_c M_{\rm nd}/L\approx
10^4(\timav/10^{-10}M_\odot\ {\rm yr^{-1}})^{-0.75}$ years for
$M=0.8M_\odot$ \citepalias{TownBild03}, so that this discrepancy
cannot be due to a temporarily low $\dot M$.  Although selection
effects favor low $\timav$'s above the period gap, since such systems
are more likely to have clean WD spectra, Patterson's (1984) estimates
of $\timav$ for the systems above the gap in Figure
\ref{fig:mdot-Porb_wide} are not systematically below his estimates
for other CVs.  The most recently improved calibration of the magnetic
braking law, using spin-down of open cluster stars \citep{Andretal03},
yielded $\timav$'s above the period gap at least a factor of ten lower
than Howell et al. (2001), falling below our inferences, so that
braking in CVs must be enhanced over that responsible for the spin
down of noninteracting low mass stars.

  In Figure 2, we also show Patterson's (1984) deduction from
observations, $\timav\approx 5.1\times 10^{-10}(P_{\rm orb}/4\ \rm hr)
M_\odot$ yr$^{-1}$, for $M=0.6M_\odot$ (dotted line) and $1.0M_\odot$
(dot-dashed line).  Our points are consistent with \citet{Patt84}
above the period gap, within the uncertainty in his estimates. Below
the gap, however, our measurements are roughly a factor of 3 above
his. The \citet{Patt84} estimates also suffer from the absence of
reliable distances, but in a more direct way than ours, making
systematic errors difficult to quantify.

\begin{figure}
\plotone{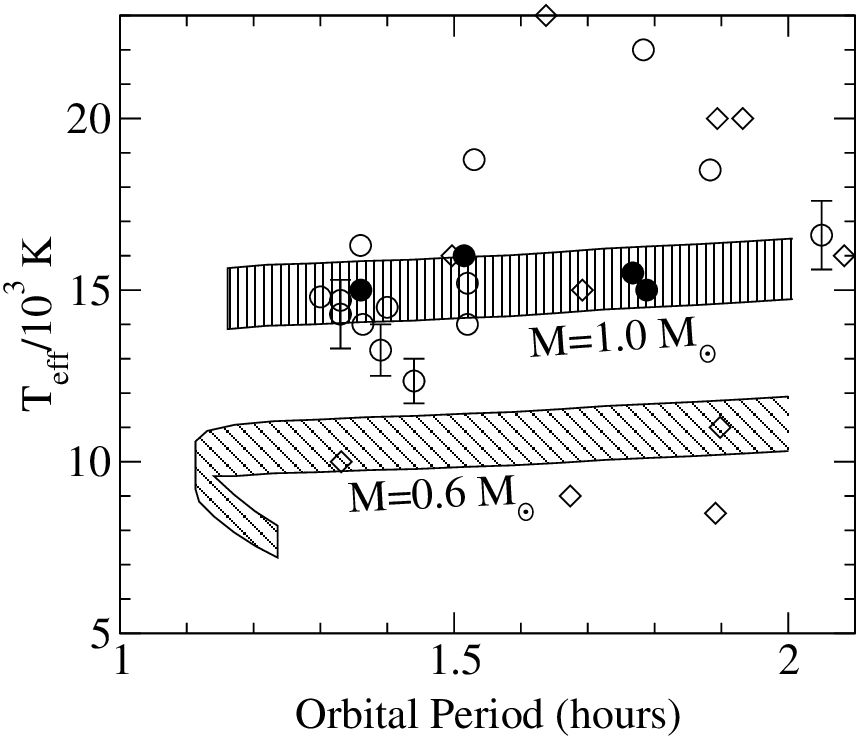}
\caption{
\label{fig:Teff-Porb}
Comparison of our predicted ranges for $T_{\rm eff}$ with observed
values for systems with $P_{\rm orb} < 2$ hours.  The filled areas
indicate for each $P_{\rm orb}$ the range of $T_{\rm eff}$ that a
quiescent CV primary is expected to traverse between thermonuclear
outbursts, using the $\timav$ expected from angular momentum loss due
to gravitational radiation \citep{KolbBara99}.  Measurements are from
the table in \citet{WintSion03} except as noted in the text.  Error
bars indicate systematic errors due to the unknown WD mass, and
represent the range of $T_{\rm eff}$ obtained with
$M=0.3$-$0.9M_\odot$.  All open points are subject to this same
uncertainty.  The filled points are systems which have well measured
WD masses (Patterson 2001), in order of increasing $P_{\rm orb}$ and
in units of $M_\odot$ the masses are $0.9\pm 0.15$, $0.82\pm 0.05$,
$0.61\pm 0.04$, and $0.84\pm 0.09$ for WZ Sge, OY Car, HT Cas, and Z
Cha.
The diamonds are magnetic CVs (\citealt{Sion99}; \citealt{GansSchm01};
\citealt{BellHowe03}).
}
\vspace{-0.3in}
\end{figure}

The best quality data are for CVs below the period gap, and
\citet{HoweGansetal02} and \citet{SzkoGansetal02} provide examples of
these measurements with a discussion of how the uncertainty in the
surface gravity, $g$, affects the $T_{\rm eff}$ results. We display
our predictions for the $T_{\rm eff}$ ranges along with observed
values for $P_{\rm orb}<2$ hrs in Figure \ref{fig:Teff-Porb}.  The
$\timav$-$P_{\rm orb}$ relation expected from gravitational radiation
losses for $M=0.6M_\odot$ is from \citet{KolbBara99}, and we use the
same mass-radius relation for the donor to find $\timav$-$P_{\rm orb}$
for $M=1.0M_\odot$. This gives $\timav = 3.6\times 10^{-11}$ and
$5.1\times10^{-11}M_\odot$ yr$^{-1}$ for $M=0.6$ and $1.0M_\odot$
respectively at $P_{\rm orb}=1.5$ hours.  Due to a difference in the
donor mass-radius relation, these differ slightly from the values of
\citet{Howeetal01} shown in Figure \ref{fig:mdot-Porb_wide}, where
$\timav =4.6\times10^{-11}$ and $6\times 10^{-11}M_\odot$ yr$^{-1}$
for $M=0.6$ and $1.1M_\odot$ at $P_{\rm orb}=1.5$ hours.

The $T_{\rm eff}$ measurements shown by circles are again from the
table in \citet{WintSion03} (with the modifications discussed
earlier).  The error bars indicate systematic errors due to the
unknown WD mass, as the spectral fits cannot independently constrain
$T_{\rm eff}$ and $g$, and represent the range of $T_{\rm eff}$
obtained for $\log g=8\pm 0.5$ ($M=0.3$-$0.9M_\odot$). All spectral
measurements of these CV WD $T_{\rm eff}$'s are subject to this same
uncertainty. The data with displayed error bars are the \emph{best}
measurements, and others have similar or larger uncertainies. The
diamonds show $T_{\rm eff}$ for magnetic systems
(\citealt{Sion99}; \citealt{GansSchm01}; \citealt{BellHowe03}). The
estimated masses for the two magnetic systems at 1.9 hours are 0.5 and
$0.6M_\odot$ \citetext{\citealp{Schwetal93}; \citealp{Schmetal83}},
for the lower and higher respectively, placing their measured $T_{\rm eff}$ 
very close to that expected from our work.

When the WD mass, and thus radius, is known, $\timav$ can be directly
constrained. The filled points are systems which have relatively
secure WD masses from eclipse timing (Patterson 2001), in order of
increasing $P_{\rm orb}$ and in units of $M_\odot$ the masses are
$0.9\pm 0.15$, $0.82\pm 0.05$, $0.61\pm 0.04$, and $0.84\pm 0.09$
for WZ Sge, OY Car, HT Cas, and Z Cha.
The best $T_{\rm eff}$ measurement is that for WZ Sge
\citep{SionChenetal95} giving $\timav =
6.4^{+1.7+3.9}_{-1.7-1.4}\times10^{-11}M_\odot$ yr$^{-1}$ where the
first errors represent the unconstrained value of $M_{\rm acc}$ and
the second the uncertain mass. If the gravitational radiation
prediction of $\timav$ \citep{KolbBara99} is taken as a lower
limit, then comparison of these CVs to the predicted $T_{\rm eff}$'s
of Figure \ref{fig:Teff-Porb} yield a maximum WD mass.  For example,
none of the CVs denoted with filled circles can have $M$ in excess
of $\approx M_\odot$, since a more massive WD
would yield a higher $T_{\rm eff}$ than observed.

  If the current $\timav$-$P_{\rm orb}$ relation predicted from
gravitational radiation is correct, this comparison favors WD masses
near $0.9-1.0M_\odot$.  However, the systems with measured masses
indicate another possible interpretation, that $M\approx
0.85M_\odot$ for many systems but $\timav$ is slightly greater than
that predicted by \citet{KolbBara99}.  The fact that the lower bound
of measured values lies roughly parallel to what is expected for
post-turnaround objects provides the exciting possibility that the
turnaround is at roughly $P_{\rm orb}=1.3$ hours and the objects with
the lowest $T_{\rm eff}$ values, HV Vir and EG Cnc, are
post-turnaround CVs.  Under this
interpretation, our work would indicate that the $\timav$'s for these
post-turnaround objects are much higher than that expected from
current modelling of evolution under the influence of gravitational
radiation.
Such an ``extra'' angular momentum loss has been mentioned by
\citet{Patt01} as a way to understand the location of the period
minimum.

\section{Conclusions}
\label{sec:conclusion} 

 We have used our theoretical work on accreting WDs
\citep{TownBild03} to translate measurements of CV WD $T_{\rm eff}$'s
into measurements of $\langle \dot m\rangle = \timav/4\pi R^2$, the
accretion rate per unit WD surface area averaged over the thermal time
of the radiative envelope ( $\gtrsim 1000$ years). 
Since the predictions of \citetalias{TownBild03} for $L(M,\timav, T_c)$ are insensitive to
$T_c$ (except where its value can be found self-consistently) we
allowed for a range of WD masses, $0.6$-$1.2M_\odot$, and
accumulated envelope masses, $0.05M_{\rm ign}\le M_{\rm acc} \le 0.95
M_{\rm ign}$, to obtain a well-quantified uncertainty on $\langle \dot
m\rangle$ from the observed $T_{\rm eff}$.  We find evidence that
above the period gap ($P_{\rm orb}>3$ hours) DN accretion rates are
slightly overestimated by CV evolution models \citep{Howeetal01}, but
that angular momentum loss is quite enhanced compared to spin down of
isolated low-mass stars \citep{Andretal03}.  Below the period gap,
$P_{\rm orb} < 2$ hours, we find that $\timav$ is larger than that
predicted by current models of gravitational radiation losses
\citep{KolbBara99} when $M=0.6M_\odot$, indicating either
larger $M$ or higher $\timav$.

It is well known that an isolated WD will pulsate when its $T_{\rm
eff}$ is in the approximate range 11000-12000 K \citep{Bergetal95}.
While a difference in the outer atmospheric composition, H/He in an
accreting WD versus pure hydrogen in the isolated case, will shift
this range, it is likely that a similar pulsation mechanism will be
active in accreting WDs.  Our calculations indicate that accreting WDs
with $M=0.6$-$1.0M_\odot$ should be near this range when $\timav =$
few$ \times 10^{-11}M_\odot$ yr$^{-1}$.  This $\timav$ is typical of
that expected in CVs when accretion is driven by gravitational
radiation
losses, $P_{\rm orb}< 2$ hours \citep{KolbBara99}.  In fact
one system, GW Lib, has been found which does exhibit precisely this
type of variability (\citealt{VanZetal00}; \citealt{SzkoGWLib02}).
Using the interior
models developed in \citetalias{TownBild03}, we are now undertaking a
seismological study of these systems.  This offers the tantalizing
possibility of measuring the WD mass, spin and 
mass of the accumulated H/He layer. 

Since the minimum light of $\timav<10^{-10} M_\odot \ {\rm yr}^{-1}$
DN in quiescence is determined by the hot WD, we can now calculate
just how deep the large-scale optical surveys are probing into the
predicted large population of CVs with very low mass companions
($<0.1M_\odot$) \citep{Howeetal97}.
A typical survey complete to $V=20$ would find all
$0.9M_\odot$ ($0.6M_\odot$) WDs with $T_{\rm eff} > 8500$ K (7000 K)
that are within 200 pc \citep{BergWese95}, and thus all DN with
$\timav >0.8\ (1.0)\times10^{-11}M_\odot$ yr$^{-1}$, including most DN
below the period gap that have not yet reached the period minimum.
Selection of the faint CVs is still a challenge;
color-selected surveys utilize their unusual combination of a hot
WD plus a main sequence M star (see \citealt{TownBild02} for an
earlier discussion and application to CVs in Galactic Globular
Clusters).
The second round of discoveries of faint CV's from the Sloan Digital
Sky Survey (SDSS), sensitive to $\sim$20th magnitude, have just been
announced \citep{Szkoetal03}.  Based on the initial set of 19 CV's and
their orbital period distribution, \citet{Szkoetal02} claim that a
large fraction of the 400 eventual CV discoveries will be of the low
$\timav$ variety below the period gap. \citet{Marsetal02} have also
reported their discovery of three such systems in the 2dF survey,
sensitive to $\sim$21st magnitude, and expect to have about 20 low
$\timav$ systems when the survey is complete.

 We thank Ed Sion for numerous conversations, Paula Szkody for
up to date information on the DN observations, and Boris G\"ansicke and
Steve Howell for comments on the manuscript.
This research was supported by the 
National Science Foundation under Grants
PHY99-07949 and AST02-05956. Support for this work was provided by
NASA through grant AR-09517.01-A from STScI, which is operated by
AURA, Inc., under NASA contract NAS5-26555. D. T. was an NSF Graduate
Fellow and L. B. is a Cottrell Scholar of the Research Corporation.



\end{document}